# Comment on "Table-like magnetocaloric effect and enhanced refrigerant capacity in $Eu_8Ga_{16}Ge_{30}$-EuO composite materials" [Appl. Phys. Lett. 99, 162513 (2011)]

A. M. G. Carvalho[1]

[1] *Laboratório Nacional de Luz Síncrotron (LNLS), Centro Nacional de Pesquisa em Energia e Materiais (CNPEM), CEP 13083-970, Campinas, São Paulo, Brazil.*

Few years ago, Chaturvedi *et al.*[1] reported their results about refrigerant capacity in $Eu_8Ga_{16}Ge_{30}$-EuO composite materials and compared one of them with other materials previously reported. However, I respectfully suggest that a few values presented in Fig. 3b of Ref. 1 are not correct.

As refrigerant capacity, Chaturvedi *et al.*[1] used the following definition:

$$\text{"refrigerant capacity"} = [-\Delta S_T]_{max} \times \delta T_{FWHM}, \qquad (1)$$

which is actually defined as the *Relative Cooling Power* (*RCP*) in Ref. 2. $[-\Delta S_T]_{max}$ is the maximum value of $-\Delta S_T$ curve and $\delta T_{FWHM}$ is its full width at half maximum. In Ref. 2, *Refrigerant Capacity* is defined as:

$$RC = -\int_{T_1}^{T_2} \Delta S_T(T, \Delta H) dT, \qquad (2)$$

which indicates how much heat can be transferred from the cold end (at $T_1$) to the hot end (at $T_2$) of the refrigerator in one ideal thermodynamic cycle. When $T_1$ and $T_2$ are the extreme temperatures of $\delta T_{FWHM}$, i.e., $\delta T_{FWHM} = T_2 - T_1$, $\dfrac{RCP}{RC} \cong \dfrac{4}{3}$ for most second-order-transition materials. It means that we can only compare *RCP* values or *RC* values, never mixing them. In other words, it is not fair or appropriate to compare *RCP* with *RC*, for different materials, calculated in the same temperature range, since *RCP* is always larger than *RC* around $[-\Delta S_T]_{max}$.

I have analyzed ten of the materials shown in Fig. 3b of Ref. 1, including the $Eu_8Ga_{16}Ge_{30}$-EuO composite (Fig. 1). Fig. 1 shows the *RCP* values reported in Ref. 1

(full symbols) and *RCP* values I have calculated from original data cited in Ref. 1 (thin edge symbols). Besides, as an extra comparison, I present the *RC* values calculated using Eq. (2) (thick edge symbols). The graph's abscissa represents the temperatures of the $\Delta S_T$ peaks (similar to the transition temperatures).

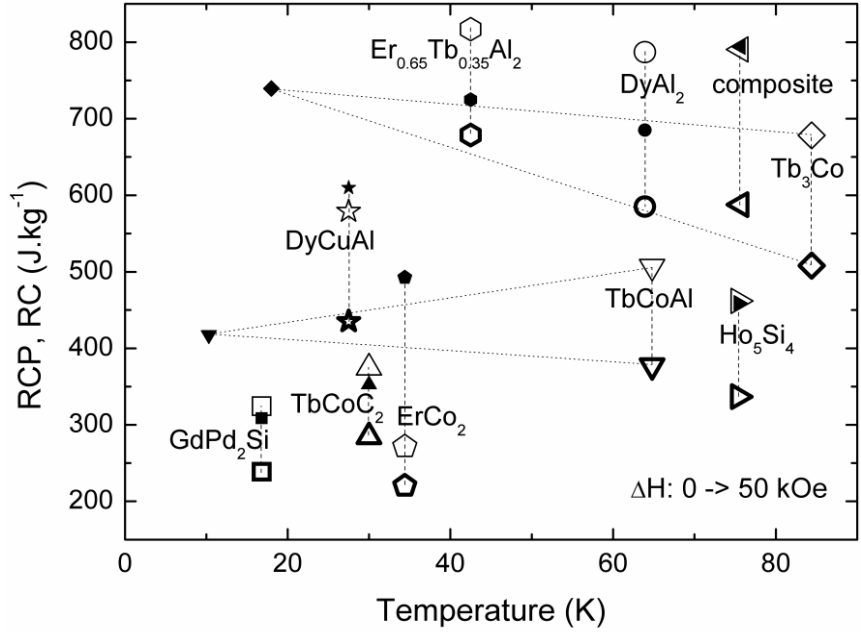

Figure 1: *Relative Cooling Power* (*RCP*) and *Refrigerant Capacity* (*RC*) values for a few materials. Abscissa represents the temperatures of the $\Delta S_T$ peaks. Solid symbols are results from Ref. 1. Thin edge symbols are the present *RCP* values. Thick edge symbols are the *RC* values. Dot and dash lines are guides for the eyes.

For $DyAl_2$ compound, I have not been able to find the Ref. 23 cited in Ref. 1, then I have used data reported by von Ranke et al.[3]. The *RCP* for $DyAl_2$ (787 J.kg$^{-1}$) is comparable to the *RCP* for the $Eu_8Ga_{16}Ge_{30}$-EuO composite (794 J.kg$^{-1}$). For $TbCoC_2$ compound, Chaturvedi et al.[1] used the *RC* value reported by Li et al.[4] (354 J.kg$^{-1}$), which should not be compared with *RCP* values of other compounds. For $Tb_3Co$ and TbCoAl compounds, besides the difference between the reported *RCP* values in Ref. 1 (739 and 417 J.kg$^{-1}$, respectively) and those I have calculated (678 and 507 J.kg$^{-1}$, respectively), the temperatures reported in Ref. 1 (18.0 and 10.3 K, respectively) are totally divergent from the temperatures of the $\Delta S_T$ peaks of these compounds (84.4 and 64.8 K, respectively). For $Er_{0.65}Tb_{0.35}Al_2$ compound, the new *RCP* value (817 J.kg$^{-1}$) is higher than that reported in Ref. 1 (725 J.kg$^{-1}$) and also higher than the *RCP* value for the $Eu_8Ga_{16}Ge_{30}$-EuO composite (794 J.kg$^{-1}$).

Summarizing, only two *RCP* values that I have calculated are coincident with those from Ref. 1 ($Eu_8Ga_{16}Ge_{30}$-EuO composite and $Ho_5Si_4$). Besides, the $Eu_8Ga_{16}Ge_{30}$-EuO composite does not have the highest *RCP* value (nor *RC* value) among the analyzed materials. Nevertheless, this is still a good candidate material for active magnetic refrigeration around the boiling point of nitrogen. I kindly suggest that Chaturvedi *et al.*[1] check the *RCP* values for the other materials presented in Fig. 3b of Ref. 1. Finally, it is not fair or appropriate to compare *RCP* values with *RC* values, for different materials, calculated in the same temperature range.

References

[1] A. Chaturvedi, S. Stefanoski, M.-H. Phan, G. S. Nolas and H. Srikanth, Appl. Phys. Lett. **99**, 162513 (2011).

[2] K. A. Gschneidner, Jr. and V. K. Pecharsky, Annu. Rev. Mater. Sci. **30**, 387–429 (2000).

[3] P. J. von Ranke, V. K. Pecharsky and K. A. Gschneidner, Jr., Phys. Rev. B **58** (18), 12110 (1998).

[4] B. Li, W. J. Hu, X. G. Liu, F. Yang, W. J. Ren, X. G. Zhao and Z. D. Zhang, Appl. Phys. Lett. **92**, 242508 (2008).